\documentclass[12pt]{iopart}
\usepackage[pdftex]{graphicx}
\begin{document}
\title[]
{Evolution of Asymmetric Raman line-shape from nano-structures}
\author{Rajesh Kumar$^{1,2,3}$ \footnote{Corresponding Author: rajeshkumar@iiti.ac.in},  Gayatri Sahu $^1$, Shailendra K. Saxena$^1$, Hari M. Rai$^1$ and Pankaj R. Sagdeo$^{1,2}$}
\address{$^1$Discipline of Physics, School of Basic Sciences, Indian Institute of Technology Indore, Madhya Pradesh-452017, India}
\address{$^2$Disciplne of Surface Science and Engineering, Indian Institute of Technology Indore, Madhya Pradesh-452017, India}
\address{$^3$Discipline of Bioscience and Bioengineering, Indian Institute of Technology Indore, Madhya Pradesh-452017, India}

\begin{abstract}
A step-by-step evolution of an asymmetric Raman line-shape function from a Lorentzian line-shape is presented here for low dimensional semiconductors. The evolution reported here is based on the phonon confinement model which is successfully used in literature to explain the asymmetric Raman line-shape from semiconductor nano-structures. Physical significance of different terms in the theoretical asymmetric Raman line-shape has been explained here.  Better understanding of theoretical reasoning behind each term allows one to use the theoretical Raman line-shape without going into details of theory from first principle. This will enable one to empirically derive a theoretical Raman line-shape function for any material if information about its phonon dispersion, size dependence etc is known. 

\end{abstract}

\maketitle

\section{Introduction}

Raman spectroscopy, since its discovery in 1928, is a widely used technique to investigate the vibrational and electronic  properties of  materials \cite{raman,raman2}. The crystalline, amorphous or nano-crystalline nature of any material can be ascertained by studying the first order Raman spectrum. The physics behind Raman scattering in semiconductors or crystals is based on the inelastic interaction of incidence photons with lattice vibrations i.e. phonons, that are sensitive to internal or external perturbations. In crystalline materials, the Raman scattering is limited to near zone centered phonons (ZCP). But in the case of nano-structures the phonons other than the zone centered ones also contribute due to confinement or localization of phonons in a finite dimension, resulted in a change in the line-shape of the first order Raman spectrum. The Raman spectrum provides a fast and convenient method to analyze the vibrational properties of crystalline and amorphous semiconductors. Since nano-structures are quasi-crystalline, their Raman spectra are expected to be intermediate between the spectra of the corresponding crystalline and amorphous materials. Many research investigations have been carried out in order to understand the vibrational properties in semiconductor nano-structures \cite{ritcher,campbell}, confined in one or more dimensions. All the results, show a shift in the peak position of the first-order Raman line of crystalline semiconductors towards lower energy accompanied by a spectral broadening and phonon softening. Although it is applicable to all the materials, in the present study, we have done a detail analysis of first-order Raman line-shape of semiconductor material especially Silicon. 

Raman scattering study on Si nano-structures has been studied extensively \cite{silicon1,silicon2,nrl3-105,nanoGS,mrs-proceeding}. Raman spectrum from the c-Si is generally symmetric having a Lorentzian line-shape appeared at 521 cm$^{-1}$ with a natural line-width of $\sim$ 4 cm$^{-1}$. Any deviation from symmetry or change in its full width at half maximum (FWHM) can be explained in terms of quantum confinement, heating related effects and electron-phonon interactions etc. More than one effect can contribute to the Raman line-shape asymmetry or broadening in the Si nano-structures. It is important to explain its behavior as a response to different perturbations present in the system. By providing an appropriate theoretical explanation, it is possible to investigate many processes taking place at the nano-scale level in the Si nano-structures. In the Si nano-structures, frequency of the optical phonons is expected to change due to finite size of the material. Thus, the Raman spectra from the Si nano-structures are also modified. This is mainly because, the translational symmetry of the crystalline materials is broken at grain boundaries, which results in the appearance of specific surface and interface vibrational contributions \cite{gouadec}. This is more pronounced in reduced dimensional systems. These properties change the frequency of the phonons in the nanostructures, which is reflected in Raman scattering. Different models are presented to depict the Raman line-shape resulting from the nano-structures. One of the well accepted model is phonon confinement model (PCM) as proposed by Richter et al. \cite{ritcher}, which is further modified by Campbell et al.\cite{campbell}.  Intensity of the first-order Raman scattering, I ($\omega$), can be written as \cite{campbell}
\begin{equation}
I(\omega)~=~\displaystyle\int^1_0\frac{exp(-q^2L^2/4a^2)}{[\omega~-~\omega(q)]^2~+~(\gamma/2)^2} d^nq,
\end{equation}
where $q$ is expressed in units of $2\pi/a$, $a$ being the lattice constant, 0.543 nm. The parameter $L$ stands for the average size (diameter) of the nano-crystals. $\gamma$ being the linewidth of the Si optical phonon in bulk c-Si ($\sim$ 4 cm$^{-1}$). Following Tubino {\it et al.} \cite{tubino}, the dispersion $\omega$(q) of the optical phonon in a spherical Si nanocrystal can be taken as 
\begin{equation}
\omega^2(q)~=~A~+~B~cos(\pi q/2),
\end{equation}
where $A~=~1.714 \times 10^{5}$ cm$^{-2}$ and $B~=~1.000 \times 10^5$ cm$^{-2}$. 
                               
In the present study, we have discussed in depth, the evolution of asymmetric Raman line-shape in nano-materials starting from a symmetric Raman line-shape from its bulk counterpart. Understanding the each term in the asymmetric Raman line-shape (due to phonon confinement effect) will enable the researchers to explore the possibility of PCM beyond elemental semiconductor nano-materials like Si, Ge and apply the same to other compound semiconductors like $GaAs$, $TiO_2$, etc. 

\section{Analysis and Discussion}

\begin{figure}[h]
\begin{center}
\includegraphics[height=6.0cm]{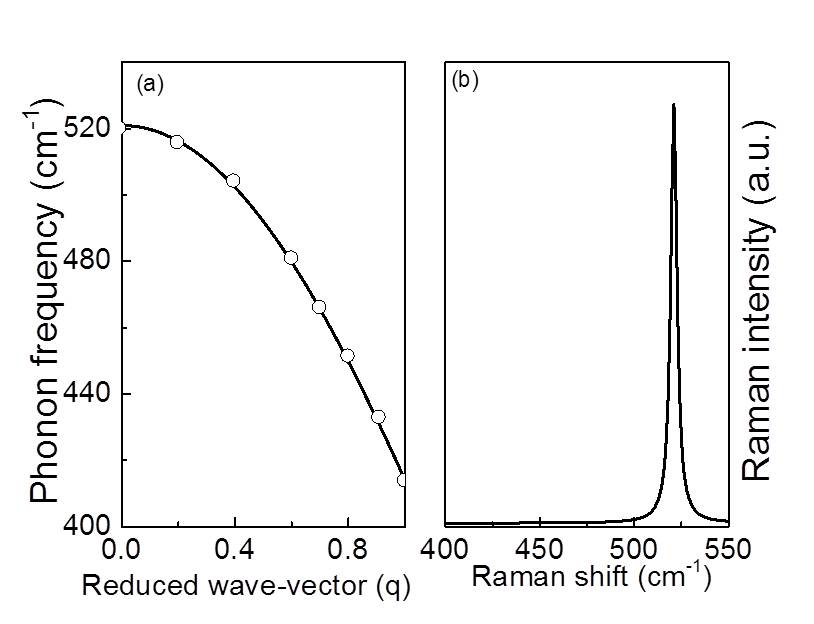}
\caption{(a) Longitudinal optic (LO) branch of phonon dispersion relation for silicon; (b) a typical Raman line-shape from bulk Si represented by Eq. 3.}
\label{raman1} 
\end{center}
\end{figure}

As discussed earlier, Raman line-shape is a manifestation of Raman active phonon mode of a semiconductor obtained after scattering of light \cite{gouadec,bergman}. It is an established fact that a Lorentzian Raman line-shape is observed from a bulk crystalline semiconductor, which is symmetric \cite{silicon2,gouadec,mavi1,mavi2} in nature whereas that from its nano counterpart is asymmetric \cite{pivac,li,wang,prusty,konstantinovic,pisc1,pisc2}. Peak of the Raman line-shape from a bulk semiconductor corresponds to the zone-center optic phonon. Figure 1(a) shows the longitudinal optic (LO) branch of phonon dispersion relation for silicon (Si). The discrete points in the figure shows the data reported in literature for the same \cite{tubino, brockhouse}. These data points are represented theoretically by Eq.  2 \cite{kumar,kumar2,shukla}.
                                  
Figure 1(b) shows a typical Raman line-shape from bulk Si represented by Eq. 3.
\begin{equation}
I(\omega)= \frac{1}{[\omega-\omega(0) ]^2+(\gamma/2)^2}    ,                                                           
\end{equation}
where, $\omega(0)$ is the zone center optical phonon frequency corresponding to q = 0 in Figure 1a and ‘$\gamma$’ is the FWHM of the Raman peak. For bulk Si, the values of $\omega(0)$ = 521 cm$^{-1}$ and $\gamma$ = 4 cm$^{-1}$, are used  to calculate the Raman line-shape as shown in Figure 1b.

Raman line-shape from nano-structured Si as given in Eq. 1 was derived by Richter et al \cite{ritcher} and later modified by Campbell et al \cite{campbell}. All the terms in Eq.1 represents a concept and is present with an underlying physics which will be discussed now. At nano-scale, phonons that participate in Raman scattering are confined within the boundary of nano-structure. Due to this quantum confinement effect breakdown of Raman selection rule takes place and phonons, away from the zone center ($q \neq 0$), also take part in Raman scattering \cite{ritcher,campbell}. As a result, non zero Raman intensities, $I(\omega)$, are  also observed at lower frequencies (in wave-number units) away from $\omega(0)$. The final Raman line-shape will be summation of all the intensities corresponding to wave-numbers at non-zero wave-vectors. After inclusion of this fact in Eq. 3, Raman line-shape function can be written as Eq. 4 as follows:
\begin{equation}
I(\omega)= \sum_q \frac{1}{[\omega-\omega(q) ]^2+(\gamma/2)^2}   ,                                              
\end{equation}
where summation over different values of ‘$q$’ (varying from 0 to 1) represents the contributions from all phonons. Since Eq. 4 involves a summation, variation in ‘$q$’ (0 to 1) can be done in different steps. Raman line-shape in Eq. 4 is plotted in Figure 2 for three different step sizes $viz.$ 0.1, 0.05 and 0.02. It is important to mention here that the plots in Figure 2 are not actual Raman line-shapes.  Spikes in Figure 2 correspond to the contribution in line-shape from individual phonon corresponding to a particular value of ‘$q$’. For example, the peak marked with asterisk (*), centered at  $\sim$ 414 cm$^{-1}$, represents the contribution of phonon represented by $q = 1$. For smaller step sizes, the line-shape is continuous instead of spiky as can be seen in Figure 2(c). Most visible feature of line-shape in Figure 2(c) is, its asymmetry as compared to the line-shape in Figure 1(b). A careful observation of evolution from Figure 2(a) - 2(c) reveals that the asymmetric line-shape arises if all phonons of a branch in phonon dispersion relation take part in the scattering. In addition to this it is also clear from Figure 2 that spikes are prominent in the lower wave-number region (400 - 475 cm$^{-1}$) as compared to the higher wave-number region (475 - 521 cm$^{-1}$). This observation is directly related to the slope of the phonon dispersion relation in Figure 1(a). The phonon dispersion relation is nearly flat near the zone center ($q = 0$) whereas it is monotonically decreasing near the zone edge ($q = 1$) making the dispersion relation nonlinear. Due to this nonlinearity, contributions from the phonons in the vicinity of zone center add together to result in higher intensity near 521 cm$^{-1}$ (corresponding to $q=0$ phonon frequency). Line-shapes in Figure 2 have been plotted by assuming that all the phonons from q = 0 to 1 contribute equally to the Raman intensity and the phonons are discrete (as represented by steps in the value of q). The actual Raman line-shape from nano-structures can now be derived by taking into account the actual contribution from all the phonons and by considering that the phonon spectrum is continuous rather than discrete \cite{campbell}.

\begin{figure}[h]
\begin{center}
\includegraphics[height=6.0cm]{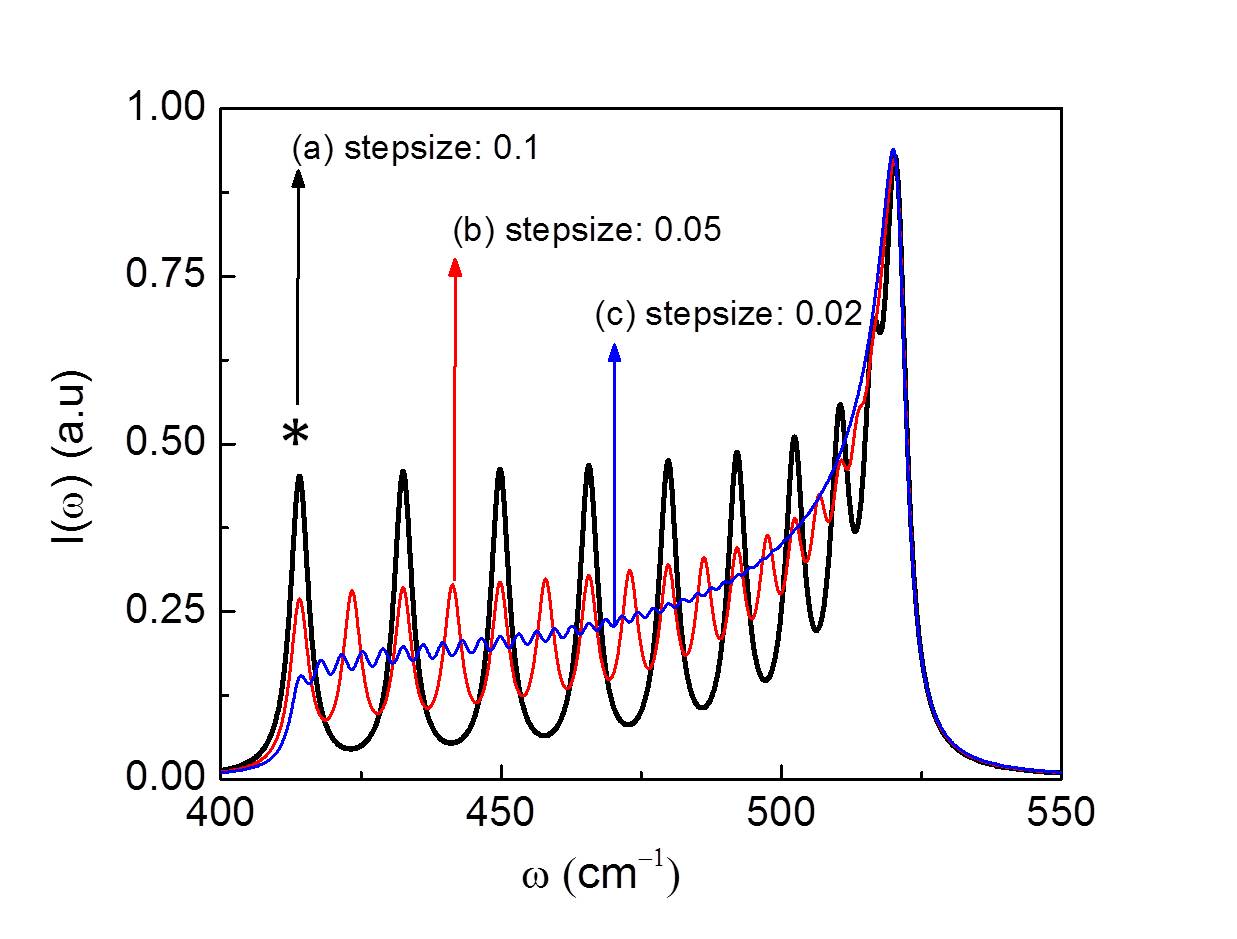}
\caption{Variation of Raman line-shape intensity as a function of wave-number plotted using Eq. 4. by considering contributions from longitudinal optic phonons at different step sizes on the dispersion curve in Fig 1(a) }
\label{raman} 
\end{center}
\end{figure}

As discussed above, the actual asymmetric Raman line-shape (as observed from nano-structures) will be derived here for the case of Si nano-structures. The contribution of the whole phonon spectrum in Raman line-shape equation can be incorporated in Eq. 4  by integrating over the full dispersion relation rather than doing summation. As a result, the resultant line-shape equation can be written as Eq. 5 as follows.
\begin{equation}
I(\omega)= \int_0^1 \frac{1}{[\omega-\omega(q) ]^2+(\gamma/2)^2} dq  ,                                              
\end{equation}

Equation 5 is plotted in Figure 3(a) no spikes are seen in the asymmetric plot due to integration over all the possible values of ‘$q$’. Almost flat intensity in the 420-480 cm$^{-1}$ region is due to the hypothesis that all phonons contribute equally or having equal probability to participate in the scattering. The actual contribution of a phonon depends on the extent of confinement inside a particular nano-crystal. This is expresses quantitatively by introducing a “weighting function” \cite{campbell} to care of the actual contribution of a particular phonon. The nature of this weighting function depends on the dimensionality of confinement. For simplicity, an exponential weighting function has been suggested by Campbell et al \cite{campbell} for all practical purposes. After introduction of the size dependence, the Raman line-shape is represented by Eq. 6 as follows:

\begin{equation}
I(\omega,L)= \int_0^1 \frac{exp.(-q^2 L^2/4a^2)}{[\omega-\omega(q) ]^2+(\gamma/2)^2} dq 
\end{equation}

Where ‘$L$’ is the size of nano-structure and ‘$a$’ is the lattice parameter of semiconductor crystal. Eq. 6 is plotted by taking the value of $L$ = 2.5 nm and $a$ = 0.543 nm (lattice constant of Si) in Figure 3(b) alongside 3(a) for comparison. By incorporating the size factor, the line-shape reflects the true asymmetric behavior as observed in the Raman spectra from nano-structures.  Incorporation of the exponential term in Eq. 6 also takes care of the size dependence of the Raman line-shape, which is in consonance with the observed size dependent Raman scattering \cite{silicon1,nrl3-105,adu1,adu2}. Equation 6 can be understood as the basic Raman line-shape for nano-structured semiconductors in particular and nano-structured materials in general. 

Modifications can be done Eq. 6 to obtain various special Raman line-shapes reported to investigate the effects of stress, temperature, electron-phonon (Fano) effect, quantum confinement effect etc. Details about these functions are available in the literature. Temperature dependence of phonon dispersion relation can directly be used if one wants to investigates the temperature dependent quantum confinement effects. In addition, degree of confinement can also be considered by using the appropriate integral (surface or volume integral) and will result in Eq. 1 from where the discussion of this paper started. If in a sample all the nano-structures are not of same dimension, a distribution function can successfully be introduced not only to take care of the distribution effect but also to know the exact size distribution available in the sample by theoretical fitting of the experimental data with the modified Raman line-shape.

\begin{figure}
\begin{center}
\includegraphics[height=6.0cm]{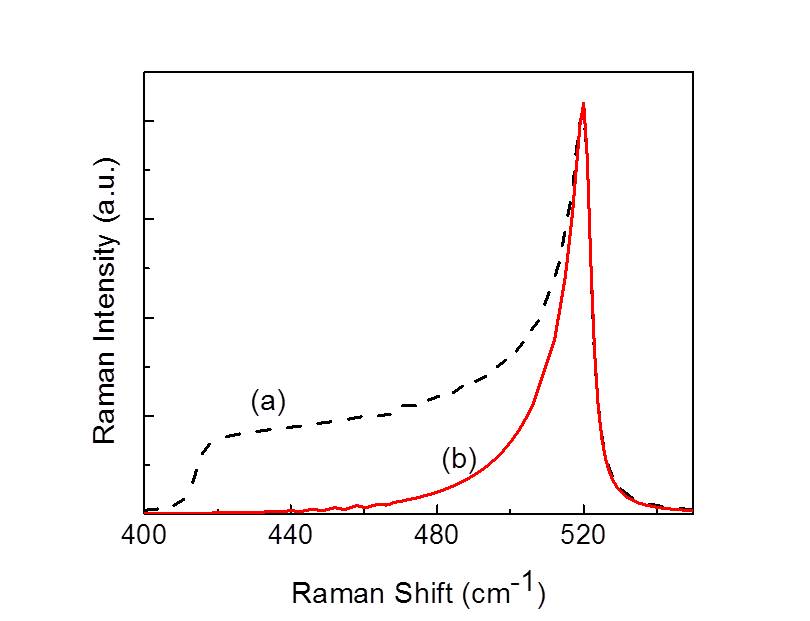}
\caption{Theoretical Raman line-shape function obtained by considering that all the phonons taking part in Raman scattering. Effect of size is not considered in (a) whereas effect of size is considered in (b).}
\label{tem} 
\end{center}
\end{figure}

\section{Summary and Conclusions}
An asymmetric Raman line-shape is observed from semiconductor nano-structures in contrast with a symmetric one from its bulk counterpart. This asymmetric line-shape is used to get information about the shape, size and distribution of nano-structures by theoretically fitting with a theoretical Raman line-shape function. Physical significance of different terms in the theoretical Raman line-shape can be explained for better consonance between theoretical Raman line-shape and experimental Raman scattering data.  Better understanding of theoretical background makes the use of Raman line-shape function more versatile and can be used for other semiconductors and materials. If information about the phonon dispersion, size dependence etc is known a Raman line-shape can be derived theoretically from the basic Lorentzian line-shape. Effects of other perturbations (like stress, temperature etc) can also be easily incorporated to get a new Raman line-shape.  

\section{Acknowledgements}
Help from Mr Ashish Kumar Yadav is highly acknowledged. One of the author (GS) would also like to acknowledge DST, Govt. of India, for the funding under DST Fast Track Scheme for Young Scientists, Project No. SR/FTP/PS-007/2012.

\end{document}